\documentclass[12pt]{iopart}
\begin{document}
\begin{flushright}
USC-00/005, CITUSC/00-063\\
\hfill  hep-th/0011207
\end{flushright}

\title[Holographic R.G. Flows] 
{Holographic Renormalization Group Flows: \newline 
The View from Ten Dimensions 
\footnote[1]{Talk presented at the  G\"ursey Memmorial
Conference II,  on {\it M-theory and Dualities,}
held at Bogazici University, Istanbul, Turkey,
June 19--23, 2000.} }

\author{Nicholas P. Warner}

\address{Department of Physics and Astronomy,  
University of Southern California \\
Los Angeles, CA 90089-0484, USA}
\address{{\it and}}
\address{CIT-USC Center for Theoretical Physics \\
University of Southern California, 
Los Angeles, CA 90089-0484, USA }

\begin{abstract}
The holographic description of supersymmetric RG flows in supergravity
is considered from both the five-dimensional and ten-dimensional
perspectives.  An ${\cal N}=1^*$ flow of ${\cal N}=4$ super-Yang Mills
is considered in detail, and the infra-red limit is studied in terms
of  IIB supergravity in ten dimensions.  Depending on the vevs
and the direction of approach to the core, the supergravity solution
can be interpreted in terms of either $5$-branes or $7$-branes.
Generally, it  is shown that it is essential to use the ten-dimensional 
description in order to study the infra-red asymptotics in supergravity.   
\end{abstract}




\def\cN{{\cal N}}
\def\cP{{\cal P}}
\def\cQ{{\cal Q}}
\def\cR{{\cal R}}
\def\cS{{\cal S}}
\def\cU{{\cal U}}
\def\cV{{\cal V}}
\def\coeff#1#2{\relax{\textstyle {#1 \over #2}}\displaystyle}
\def\neql#1{{\cal N} \! =\! #1}  
\def\half{{1 \over 2}}
\def\IR{\relax{\rm I\kern-.18em R}}
\def\IP{\relax{\rm I\kern-.18em P}}
\font\cmss=cmss10 \font\cmsss=cmss10 at 7pt
\def\ZZ{\relax\ifmmode\mathchoice
{\hbox{\cmss Z\kern-.4em Z}}{\hbox{\cmss Z\kern-.4em Z}}
{\lower.9pt\hbox{\cmsss Z\kern-.4em Z}}
{\lower1.2pt\hbox{\cmsss Z\kern-.4em Z}}\else{\cmss Z\kern-.4em
Z}\fi}
\def\eop{\mathop{e}^{\!\circ}{}}
\def\gop{\mathop{g}^{\!\circ}{}}
%
\def\atmp#1{Adv.\ Theor.\ Math.\ Phys.\ $\underline{#1}$\  }
\def\jhep#1({J.\ High \ Energy \ Phys. $\underline {\bf #1}$\ (}
\def\mnup#1{Nucl.\ Phys.\ {\fbbx B#1}\ }
\def\nup#1{Nucl.\ Phys.\  $\underline {B#1}$\ }
\def\plt#1{Phys.\ Lett.\ $\underline  {B#1}$\ }
\def\cqg#1{Class.\ Quant.\ Grav. \ $\underline {B#1}$\ }
\def\cmp#1{Comm.\ Math.\ Phys.\ $\underline  {#1}$\ }
\def\prp#1{Phys.\ Rep.\  }
\def\prl#1{Phys.\ Rev.\ Lett.\ $\underline  {#1}$\ }
\def\prv#1{Phys.\ Rev.\ $\us  {#1}$\ }
\def\mpl#1{Mod.\ Phys.\ Let.\ $\underline  {A#1}$\ }
\font\ninerm=cmr9
\font\ninebf=cmbx9
\font\ninesl=cmsl9 
\def\nihil#1{{\it #1}}
\def\eprt#1{{\tt #1}}
 

\section{Introduction}

Since its original formulation, the holographic AdS/CFT correspondence  
\cite{JMalda,HolDims} has undergone some remarkable developments
and extensions.  In particular the holographic correspondence has been
generalized to many non-conformal settings, and most particularly in
the study of renormalization group (RG) flows.  For the latter
\cite{flowpaps,FGPWa} one starts from a UV conformal fixed point field
theory on a brane, represented by an AdS solution in supergravity.
Fields on the brane then represent boundary data (at infinity in AdS)
for a supergravity solution \cite{HolDims}:  That is, one  perturbs the
theory on the brane and finds the corresponding supergravity solution 
with the boundary data at infinity determined by the perturbing fields.  
The renormalization group flow is then believed to be
given by looking at the radial evolution of the supergravity solution
in a direction transverse to the brane.  The cosmological
scale factor in front of the brane metric becomes the RG scale, and a
generic, positive energy perturbation on the brane causes this
cosmological factor to decrease with the radius \cite{FGPWa}.  Thus the deep
interior of the supergravity solution probes the infra-red structure
of the perturbed field theory.

In this paper I will consider supersymmetric flows generated by mass 
perturbations, and special vevs in  $\cN=4$ Yang-Mills, and in particular 
I will focus on the infra-red structure suggested by the dual supergravity
solutions.  There are several reasons for considering these
particular flows.  First,  the duality between IIB supergravity and 
large $N$, $\cN=4$ Yang-Mills on $D3$-branes is perhaps
the most extensively studied holographic correspondence and is therefore a 
natural starting point for studying non-conformal generalizations of 
the holographic correspondence.   Secondly, supersymmetric flows afford 
the best oppurtunity to
test the non-conformal, holographic correspondence.  This is primarily
because the supersymmetry, and associated non-renormalization
theorems, enable the exact computation of some parts of the quantum
effective action, or the determination of exact beta-functions \cite{NSVZ}
in the field theory on the brane.  Moreover, the question of whether a
supergravity calculation for large $N$ field theory has a well
defined, finite $N$ counterpart is essentially the same as whether a
solution to the supergravity equations represents a background
``vacuum'' for the complete string theory.  One of the
lessons of the last fifteen years has been that supersymmetric
supergravity solutions can usually be generalized to string theory.
In holographic language this means that supersymmetric supergravity
solutions have a good chance of telling us something about finite $N$
field theories, while non-supersymmetric solutions are in danger of
merely probing large $N$ pathologies.

While the $D3$-brane/IIB supergravity correspondence is framed in 
ten dimensions, it is possible to reduce one of the most important
sectors down to a five-dimensional description.  More precisely,
gauged $\cN=8$ supergravity in five dimensions provides an exact
description of the ``massless'' sector of IIB supergravity reduced on
$S^5$. In terms of the physics on the brane, this means that one can
use $\cN=8$ supergravity in five dimensions to study mass
perturbations and special vevs in $\cN=4$ supersymmetric Yang-Mills
theory.  The benefit of using the five-dimensional formalism is that
the structure and equations of motion are far simpler than their
ten-dimensional counterparts, and this has enabled extensive analysis
of RG flows, and most particularly for the supersymmetric flows (see,
for example \cite{ KPW, FGPWa, FGPWb, GPPZ, Beh, PetZaf, NEMP}). 

The reduction of the supergravity flow to a five-dimensional
problem thus provides an extremely powerful computational tool,
but there is a price:  the correct physical interpretation
of the flows is far from transparent in the five-dimensional
world.  Indeed, it is very easy to arrive at the wrong physical
picture because of this five-dimensional ``myopia''.  It is
one of the primary purposes of this paper to show that 
the proper physical interpretation of the supergravity solutions 
{\it requires} that one  find their ten-dimensional counterparts.  
It is important to stress
that the five-dimensional solutions are neither wrong, nor incomplete,
but that they encode the ten-dimensional data in a rather subtle
manner that can obscure much of the physics.  I will illustrate
this in some detail for the holographic description of 
the ${\cal N}=1^*$  flows in which $\cN=4$ Yang-Mills is broken to 
$\cN=1$ by giving an 
equal  mass to all the chiral multiplets \cite{GPPZ,PolStr,KPNWc}.  
However, the issues addressed here have far wider consequences, 
especially  for attempted descriptions of field theory via 
five-dimensional ``brane worlds'' (see, for example, \cite{RS,BVV}).

\section{Flows in five dimensions}

I will be considering the special class of fields in
IIB supergravity that correspond to the ``massless''
perturbations of the $AdS_5 \times S^5$ background.
As is,  by now, well known, this set of fields forms
a consistent, closed subsector of the IIB supergravity
theory, and the complete action for this subsector is
given by that of gauged $\cN=8$ supergravity
in five dimensions \cite{GaugSG}.  In particular it
is believed (with very considerable evidence) that
any field configuration  of the five-dimensional 
theory can be embedded in the ten-dimensional theory,
and that solutions to the five-dimensional equations
of motion ``lift'' uniquely to solutions of the ten-dimensional
theory.   Thus working in the five-dimensional theory
restricts the class of fields considered, but does
not restrict the validity of the solution in
ten dimensions.

The five-dimensional solutions corresponding to RG flows 
have $5$-metrics of the form:
\begin{equation}
ds^2 = e^{2 A(r)} \eta_{\mu\nu} dx^\mu dx^\nu - dr^2 \ .
\label{genmetric}
\end{equation}
(I will adopt the conventions of \cite{FGPWa} throughout.)  
This metric preserves Poincar\'e invariance on the $D3$-brane,
but allows for a ``cosmological'' scale factor, and in 
the flow interpretation, $e^{A(r)}$ represents the RG  
scale on the branes.   For the flow solutions, one requires
that $A(r) \sim r/L$ as $r \to \infty$ so that the flow
starts with the conformally invariant $AdS_5$ metric of radius $L$.
One also usually only considers
non-trivial scalar backgrounds that depend solely on $r$ in the 
five-dimensional supergravity since
this also respects the Poincar\'e invariance on the branes.
In $\cN=8$ gauged supergravity, there are forty-two scalars,
$\varphi_j$, 
and in the  Yang-Mills theory these scalars are dual to the  
gauge coupling, the $\theta$-angle, the fermion bilinear 
operators and the scalar bilinear operators.
The latter have the form:
\begin{equation}
{\rm Tr}~(\lambda^a \, \lambda^b)\ ,  \quad
{\rm Tr}~(\bar \lambda^{\bar a} \, \bar \lambda^{\bar b}) \ , \quad
{\rm Tr}~(X^A \, X^B)~-~{1 \over 6} ~\delta^{AB}~{\rm Tr}~(X^C\, X^C) \ . 
\label{bilins}
\end{equation}
In terms of representations of the $SO(6)$ $\cR$-symmetry, the
foregoing operators
transform as ${\bf  10 \oplus \overline{10}  \oplus 20'}$.

The supergravity theory has a highly non-trivial scalar
potential, $\cP(\varphi_i)$, and the scalar equations of
motion reduce to
\begin{equation}
{d^2 \over dr^2} \, \varphi_j ~+~ 4 \,A^\prime (r)\,  {d \over dr } 
\varphi_j ~-~ {\partial \cP\over \partial \varphi_j}  ~=~ 0\ . 
\label{scalareqs}
\end{equation}
For each scalar field there are generically two solutions 
to this equation, one of which represents a non-normalizable
mode of $AdS_5$, and the other representing a normalizable
mode.  The former represents the insertion of a mass term,
{\it i.e.} the insertion of one of the operators in
(\ref{bilins}) into the brane action, while the normalizable
modes represent modifications to the state on the 
brane through operator vevs.

If one requires that the flow preserve at least one
supersymmetry on the brane then one must restrict 
how the scalars flow.  In practice this is usually
done by restricting  to a subset of the scalars.
One first shows that this restriction is consistent with the
equations of motion, and then this restricted class
of flows can usually
be characterized in terms of a first order system:
\begin{equation}
{d \varphi_j \over d r} ~=~ {g \over 2}~{\partial W \over \partial 
\varphi_j} \ , \qquad {\rm and} \qquad A'(r) ~=~ - {g \over 3}~W \,,
\label{susyflows}
\end{equation}
where, on the restricted subsector, 
$W$ is a superpotential that is related to $\cP$ via:
\begin{equation}
\cP ~=~ {g^2 \over 8}~\sum_{j = 1}^3 ~\bigg| {\partial W
\over \partial \varphi_j} \bigg|^2 ~-~ {g^2 \over 3}~\big|W \big|^2 \,.
\label{VfromW}
\end{equation}
In these expressions, $g =2/L$, is the coupling of the gauged supergravity, 
and $L$ is the radius of the asymptotic $AdS_5$ near infinity.   
Henceforth I shall normalize $L$ by setting $L=1$.

One can show very generally that $A'(r)$ must increase monotonically as
$r$ decreases \cite{FGPWa}, and it follows that $e^{2 A(r)}$
will go to zero as as $r$ decreases.
There are thus two possibilities:  either $e^{2 A(r)}$
vanishes only as $r \to -\infty$, or it vanishes at finite
$r$ with
\begin{equation}
 e^{A(r)} ~\sim~ (r-r_0)^{2 \, a}\,,
\label{asymA}
\end{equation}
for some exponent $a$.  In terms of the superpotential,
it can either flow to a critical point, or it can ``flow
to Hades:'' $W \to - \infty$.
The former represents a flow to a non-trivial, supersymmetric IR fixed
point theory on the brane, and
the geometry is non-singular, and interpretation
is straightforward.  An example of this was studied
in detail in \cite{FGPWa,KPNWa}.    

Flows to Hades, in which $ e^{A}$ behaves as in (\ref{asymA}), appear 
to be singular as five-dimensional geometries, and one might be concerned 
that they are unphysical.  However, there are quite a number
of examples of physically sensible flows that have such asymptotics.
The reason why we know that such flows are sensible is that we can 
lift them to ten dimensions, and analyse them in the IIB supergravity 
theory (see, for example, \cite{FGPWb}).
I will return to this in the next section.  For the present
I simply wish to note that the five-dimensional metric 
contains rather minimal information about the IR asymptotics:
All one has is the exponent, $a$.

The flows that I particularly wish to study here are the
${\cal N}=1^*$ flows considered in \cite{GPPZ}.  On the brane the 
$\cN=4$ Yang-Mills theory is broken to $\cN=1$, large N, ``super-QCD''
by giving all three chiral multiplets a mass.  For simplicity,
these masses are taken to be equal, and the resulting theory
preserves a global $SO(3)$ symmetry.  In terms of fermion bilinears,
this means that one is turning on a (complex) supergravity scalar, $m$, 
that is dual to:
\begin{equation}
m\, \sum_{j=1}^3 \, {\rm Tr}~(\lambda^j\, \lambda^j) ~+~
\overline m\, \sum_{\bar j=1}^3 {\rm Tr}~(\bar \lambda^{\bar j}\, \bar 
\lambda^{\bar j}) \,.
\label{massterm}
\end{equation}
While there is no explicit low energy supergravity field that
is dual to the corresponding scalar mass term,
$ \sum_{A=1}^6 {\rm Tr}~(X^A X^A)$, the supergravity theory
always implicitly adjusts the solution to contain the proper amount 
of this  field as required by supersymmetry.  
One can also consider what happens if the gaugino condensate
$\sigma \equiv  {\rm Tr}~(\lambda^4 \lambda^4) $ is allowed
to run simultaneously.

Since they will play a major role later, it is important to
understand the residual $\cR$-symmetries of this scalar
sub-sector.  The original symmetry of the theory
is $SO(6) \times SL(2,\IR)$, where the latter describes
the scalar coset of the IIB supergravity.  The $SL(2,\IR)$ 
is a symmetry of the supergravity, but it is broken to
$SL(2,\ZZ)$ in the string theory.   In particular, the $U(1)
\subset SL(2,\IR)$ represents ``continuous duality symmetry''
that must be broken in the string theory.
In terms of the field theory on the brane, this ``duality'' $U(1)$
will be a property of the large $N$ theory, and will be 
broken at finite $N$.
As remarked above, the mass perturbation preserves at least
an $SO(3) \subset SO(6)$, which acts in the $SO(3)$ vector 
representation on the first three fermions $\lambda^i$,
$i=1,2,3$. This $SO(3)$ also remains unbroken when 
$\sigma$ is turned on as well. The $SO(3)$ commutes with an 
$SO(2)$ inside $SO(6)$.  Specifically, the generators of this $SO(3) 
\times SO(2)$ are the $6\times 6$ matrices:
\begin{equation}
 \left( \matrix{J_i &  0 \cr  0 & J_i } \right)  \,, \qquad
\left( \matrix{0 & {\bf I}_{3\times 3} \cr 
-{\bf I}_{3\times 3} & 0} \right)  \,.
\label{SOgens}
\end{equation}
This $SO(2)$ rotates $m$ and $\sigma$ by different phases.
Any one, but not both, of these phase rotations can be undone by a 
$U(1)$ duality rotation.  Hence turning on $m$ or $\sigma$ alone
preserves an $SO(3) \times SO(2)$\footnote{In fact, flows 
that involve $\sigma$ alone preserve an $SU(3) \times U(1)$.}, however,
this extra $SO(2)$ rotation involves a $U(1)$ duality rotation
and so will not be a symmetry of the finite $N$ field theory.

When both $m$ and $\sigma$ are turned on, the  only remaining
continuous symmetry is  $SO(3)$. However, there is still 
residual discrete symmetry:  Let $S = \left( \matrix{0 &  1 \cr  -1 &0 } 
\right) \in SL(2,\IR)$, and define $R$ to be the discrete,
$90^\circ$ rotation defined by taking the second matrix in 
(\ref{SOgens}) to be an element of the {\it group} $SO(6)$. 
The combination of these, $RS$, has order $4$, and is a symmetry
of the flows for arbitrary $m$ and $\sigma$.  Observe that
$S$ is indeed the $S$-duality transformation in 
$SL(2,\ZZ) \subset SL(2,\IR)$, and indeed this should be
a symmetry of the string theory.  It follows that whatever
structure one finds in the infra-red, one will find its 
$S$-dual at  $ 90^\circ$ to that structure.  This means
that one can never see confinement in this class of
supergravity solutions without fine tuning of the Wilson
or 't Hooft loop:  Any confining loop can always 
break to a screened loop by making a $ 90^\circ$ change
of direction near the core of the solution.

The superpotential on this sector is:
\begin{equation}
W ~=~ -{3 \over 4}\,\bigg(\,\cosh(2 \sigma) ~+~ \cosh\bigg(
{2\,m \over \sqrt{3}} \bigg) \,\bigg) \,,
\label{superpot}
\end{equation}
where I have used the symmetries of the problem to take $m$
amd $\sigma$ to be real. The equations (\ref{susyflows}) 
then have solutions \cite{GPPZ}:
\begin{eqnarray}
m &~=~  {\sqrt{3} \over 2}\,\log\bigg({ 1 + t \over  1- t } \bigg)\,,
\qquad \sigma ~=~  {1\over 2}\,\log\bigg({ 1 + \lambda t^3 \over  1- 
\lambda  t^3} \bigg)\,,  \\
 A(r) &~=~
\coeff{1}{6} \log\big( t^{-3} -\lambda^2 \, t^3 \big) ~+~
\coeff{1}{2} \log\big(t^{-1} -t\big) ~+~ C_1\,, 
\label{flowsoln}
\end{eqnarray}
where
\begin{equation}
t ~=~ \exp\big[ -\big(r - C_1\big) \big] \,, 
\qquad \lambda ~=~ \exp \big[3\big(C_2 -C_1\big)\big]  \,,
\label{paramsol}
\end{equation}
and where the $C_j$ are constants of integration for the 
flows of $m$ and $\sigma$.  As noted in \cite{GPPZ}, the physical 
metrics should have $\lambda \le 1$ since one should not
have the   gaugino condensate diverging before
the mass scale diverges.

The five dimensional metric  becomes:
\begin{equation}
ds^2 =  - dr^2 ~+~(t^{-1} - t \big)\, ( t^{-3} -\lambda^2 \, t^3 
\big)^{1 \over 3}\, \eta_{\mu\nu} dx^\mu dx^\nu\ .
\label{GPPZmetric}
\end{equation}
This is manifestly singular at $t=1$, and the exponent,
$a$, is ${1 \over 2}$ for $\lambda <1$, and ${2 \over 3}$ for 
$\lambda = 1$.  The structure of this singularity 
elucidates very little of the physics.  To go any
further, it is necessary to reconstruct the ten-dimensional
counterparts of these flows.

\section{The view from ten dimensions}

It was shown in \cite{KPW,KPNWb} that the metric and 
dilaton configuration in ten dimensions can be reconstructed
from the scalar field configuration in five dimensions.
Specifically, let $( {{\cal V}^{IJ}}_{ab},
{{\cal V}_{I\alpha}}^{ab})$ be the components of the
$E_{6(6)}$ matrix that characterizes the scalar fields
as in \cite{GaugSG}, and let  $( \widetilde{\cal V}_{IJ}^{\ \ \ ab},
 \widetilde{\cal V}^{I\alpha}_{\ \ \ ab})$ be the inverse of
${\cal V}$.   Let $K^{IJ\, p}$ be the fifteen Killing vectors
on the round $S^5$, and let $x^J$ denote the Cartesian
coordinates of the standard embedding of $S^5$ in
$\IR^6$.   Then the ten-dimensional metric has
the form
\begin{equation}
ds_{10}^2 ~=~ \Delta^{-{2 \over 3}}\, ds_{5}^2  ~+~d \tilde  s_{5}^2 \,,
\label{tenmetric}
\end{equation}
where $ds_{5}^2$ is the metric, (\ref{genmetric}), of the 
five-dimensional theory, and $ d \tilde  s_{5}^2 = \tilde g_{mn} 
dy^m \, dy^n$ is the metric on the  deformed five-sphere. 
The inverse of this metric is then given by
\begin{equation}
\Delta^{-{2 \over 3}}\,\widetilde g^{pq} ~=~ 
 {1 \over c^2}\, K^{IJ\, p}\,K^{KL\, q}\, \widetilde
\cV_{IJab}\,\widetilde
\cV_{KLcd}\, \Omega^{ac}\,\Omega^{bd} \ ,
\label{metform}
\end{equation}
where $c$ is a normalization constant, and $\Omega^{ab}$
is a $USp(8)$ symplectic form.
The warp-factor, $\Delta$, is defined by
\begin{equation}
\Delta \,\equiv\,  \sqrt{{\rm det}( g_{mp}\,\gop^{pq})} \,.
\end{equation}
where the inverse metric, $\displaystyle \gop^{pq}$, is 
that of the ``round'' $S^5$.   One can determine $\Delta$
from (\ref{metform}) by taking the determinant on both sides.

Similarly, if $\cS \in SL(2,\IR)$ describes
 the dilation/axion of the IIB theory, then the gauge independent
 quantity $\cS\cS^T$ is given by
\begin{equation}
\Delta^{-{4 \over3}}\, (\cS\,\cS^T)^{\alpha \beta} ~=~
{\rm const}\times  \,\epsilon^{\alpha \gamma} \epsilon^{\beta \delta}\,
\cV_{I \gamma}{}^{ a b} \,\cV_{J \delta}{}^{ c d} 
\,x^I x^J\, \Omega_{ac}\, \Omega_{bd} \,. 
\label{dilform}
\end{equation}

While (\ref{tenmetric}) was expected based upon earlier
results in other supergravity theories \cite{MetAns}, equation
(\ref{dilform}) was something of a surprise.  Since the 
five-dimensional potential
is invariant under the $SL(2,\IR)$ it was natural to think of
these scalars as representing the holographic duals of the
gauge coupling and $\theta$-angle.  This, however, is only correct
at the UV fixed point where the five-dimensional $SL(2,\IR)$
represents the UV gauge couplings, which are true moduli of the theory.
The ten-dimensional dilation/axion coset represents the gauge
couplings as they flow, and mix with other couplings.  The radial
dependence thus gives the running gauge couplings on the brane.

Equation (\ref{dilform}) is thus important for several reasons.
It represents an integrated version of the Callan-Symanzik
equation, and for supersymmetric flows it must also contain
an integrated NSVZ beta-function \cite{KPNWb}.
This result also represents a cautionary
tale for brane-worlds:  The relationship between the five-dimensional
variables, and the ten-dimensional fields can be quite subtle
and complicated.  In particular, the ten-dimensional dilaton and axion
can behave in a highly non-trivial manner while the five dimensional
fields in the corresponding $SL(2,\IR)$ remain fixed.  
{\it It is only by lifting to the ten-dimensional solution that
such relationships become manifest, and it is the ten-dimensional
string theory, and not the five-dimensional field theory,
 that represents the ``controlling authority'' in IIB holography.}

Returning to the ``super-QCD,'' or $\cN=1^*$,  flow described 
in the previous section,
one finds that the lift to ten dimensions has a rich structure.
To describe it in detail, I must first parametrize $S^5$ in a rather
unusual manner.   Consider $S^5$ as the unit sphere in $\IR^3
\times \IR^3$:
\begin{equation}
\vec u^2 ~+~ \vec v^2 ~=~ 1\,, \quad \vec u\,, \vec v \in \IR^3 \,.
\end{equation}
The $SO(3)$ $\cR$-symmetry invariance of the flows acts on
$S^5$ by simultaneous rotation of $\vec u$ and $\vec v$.  Using
such a rotation, we can reduce $\vec u$ and $\vec v$ to the
form:
\begin{equation}
\vec u~=~ (0,0,y_1) \,, \qquad \vec v~=~ (0,y_2,y_3)  \,,
\end{equation}
where $\sum_j y_j^2 =1$.  Thus, the $5$-sphere may be 
generically thought of an $SO(3) \equiv \IR \IP^3$ fiber
over an $S^2$ base parametrized by $\vec y$.  However,
an $SO(3)$ rotation can negate any two Cartesian coordinates,
and so the base is really $S^2/(\ZZ_2 \times \ZZ_2)$,
where we take $y_1, y_2 \ge 0$.  This represents a 
quarter of a sphere, or a disk.  Note that the perimeter
of this disk is characterized by points at which $\vec u$
and $\vec v$ are parallel.  If $\vec u$
and $\vec v$ are indeed parallel, then
the $SO(3)$ action degenerates:  there is an $SO(2)$ subgroup
that fixes $\vec u$ and $\vec v$, and the fiber is thus 
$S^2 = SO(3)/SO(2)$.  The geometric picture
of $S^5$ is thus an $SO(3)$ fibration over a unit disk, where
the fiber degenerates to an $S^2$ at the perimeter.  It
is covenient to parametrize the disk in terms of the
$SO(3)$-invariants:
\begin{equation}
 w_1 ~=~ 2 \, \vec u\cdot \vec u - 1   \,, 
\quad w_2 =  2 \, \vec u\cdot \vec v   \,,
\quad {\rm where\ } \ 0 \le w_1^2 + w_2^2 \le 1\,. 
\end{equation}

So far all I have done is find a complicated description of
$S^5$, but this description is precisely adapted to the
$\cN=1^*$ flows.

\subsection{Flows with $\lambda < 1$: Seven-branes}

Flows with $\lambda < 1$ (see  (\ref{flowsoln}),(\ref{paramsol})) 
have the gaugino condensate, $\sigma$, flowing ``slowly,'' and as we 
will see it is the chiral multiplet mass that dominates in the
infra-red, and determines the structure of the core of
the supergravity solution.  As we saw in (\ref{GPPZmetric}),
the $5$-metric becomes singular as $t \to 1$.  However, in
this limit one has $dr \sim {dt \over t}$ and 
$\Delta^{-{2 \over3}}  \sim {1 \over 1-t}$.
When substituted into (\ref{tenmetric}), the simple pole in 
$\Delta^{-{2 \over3}}$ cancels the zero in front of the
brane metric in (\ref{GPPZmetric}), and the apparent
singularity in $ {dt^2 \over t^2 (1-t)}$ can be removed
by a change of variables $\chi = (1-t)^\half$.  Finally, the
$\IR \IP^3$ fiber metric, $d \Omega_3^2$, in the $S^5$ gets
a prefactor of $(1-t) \sim \chi^2$.  Thus one finds that
the ten-dimensional metric (\ref{tenmetric}) is asymptotic to:
\begin{eqnarray}
ds^2 = & \coeff{1}{\sqrt{2}} \, {\widehat \Omega}^{1 \over 4}
~ \bigg[ 2 \,
( 1- \lambda^2)^{1/3} \,
e^{2\, C_1} \,\big(\eta_{\mu \nu} dx^\mu dx^\nu \big) - d \chi^2 -
\coeff{1}{4}\,\chi^2\, (\sigma_1^2 + \sigma_2^2  +\sigma_3^2 ) \bigg] 
\cr & ~-~ \coeff{1}{\sqrt{2}} \, {\widehat \Omega}^{-{3\over 4}}~ 
\bigg[ {2(1- \lambda) \over  (1+ \lambda)}~dw_1^2   ~+~ 
{(1+ \lambda) \over  2(1- \lambda)}~dw_2^2   \bigg]\,,
\label{sbrane}
\end{eqnarray}
where 
\begin{equation}
\widehat \Omega  ~=~ (1- w_1^2 -w_2^2) \,.
\end{equation}
In taking this limit I have assumed that $(1-t)$ is small compared
to $1- w_1^2 -w_2^2$.

Observe that this metric is regular, except on the ring
$w_1^2 + w_2^2=1$.  Moreover it consists of a flat $(7,1)$-metric
and flat Euclidean $2$-metric with warp-factors that are
precisely appropriate to the reduction of the ten-dimensional theory
down to eight dimensions.  The throat of the $D3$-brane 
metric is therefore rounding out into a $7$-brane world at $t \sim 1$,
or at $r \sim C_1$.  From (\ref{flowsoln}) we have $m \sim \sqrt{3} 
e^{C_1}  e^{-r}$ as $ r \to \infty$, and hence the value of $r$ at
which the throat rounds out is set by the logarithm of the  UV  
mass scale.   Also note that the scale on
the $D3$ brane limits to a finite value  set by the UV
mass scale ($e^{C_1}$) and by $(\lambda-1)$.  This suggests that
this family of flows cannot access the far infra-red, and that the
chiral multiplets do not fully decouple.  It also
suggests that the flows for  $\lambda=1$ should be qualitatively
different in this respect.

It is also amusing to note that the $(7,1)$-metric in (\ref{sbrane})
is actually not $\IR^{7,1}$ but $\IR^{3,1} \times  \IR^4/\ZZ_2$.
This is because $d\Omega_3^2$ is the metric on $\IR\IP^3$ and
not $S^3$.  The presence of this $A_1$ singularity might well
prove interesting in the string theory.

Finally, the dilaton background also develops a ring singularity
that supports the $7$-brane picture: 
\begin{equation}
\cS\,\cS^T ~\sim~  \widehat \Omega^{-\half} \,
\left(\matrix{1 + w_1 & w_2 \cr w_2 & 1- w_1 }\right) \,.
\end{equation}

\subsection{The ring singularity: Duality-averaged branes}

In the foregoing I considered a limit in which $(1-t)$ was
small compared to  $1- w_1^2 -w_2^2$.  It is instructive
to consider the opposite order of limits since it probes
the ring singularity more closely.  
In particular, setting 
$w_1  = \cos( 2\theta),  w_2 =   \sin (2 \theta)  \cos\phi$,  
one finds an asymptotic dilaton configuration of the form:
\begin{equation}
\cS\,\cS^T ~\sim~   Q \cdot \left(\matrix{\cU\, {1 \over  
\sqrt{1-t}}  &  0 \cr 0 & \cU^{-1} \, \sqrt{1-t}} \right) 
\cdot Q^{-1} \,,
\label{ringdil}
\end{equation}
where
\begin{equation}
\cU ~\equiv~ \Big( {2(1-\lambda^2)   \over 1+2 \,\lambda \, 
 \cos(4\,\theta) + \lambda^2 } \Big)^{1 \over 2}  \,, \qquad
Q ~\equiv~ \left(\matrix{\cos \theta & -\sin \theta 
\cr \sin \theta &\cos \theta } \right) \,.  
\end{equation} 
In terms of the IIB coupling, $\tau$  one has:
\begin{equation}
\tau ~=~ {i \,\cU\, \cos \theta ~-~ \sqrt{1-t}\, 
\sin \theta \over \sqrt{1-t} \cos  \theta ~+~
i\,\cU \,  \, \sin \theta } ~\sim~ \cot  \theta
\quad {\rm as} \ t \to 1\,.
\label{taures}
\end{equation}
Observe that the singularity structure depends only rather
weakly upon $\lambda$, and that, apart from a varying, but
positive normalization factor, the ring singularity appears
to be rotationally invariant.

Recall that for $\lambda =0$ the flow has an $SO(3) \times
SO(2)$ symmetry, and that the $SO(2)$ symmetry is a combination
of a duality rotation and the geometric rotation of $\vec u$ into
$\vec v$.  Equation (\ref{ringdil}) shows that the ring singularity 
is generated by precisely such a symmetry:  the singularity
is smeared into the ring with a simultaneous duality rotation.
Moreover (\ref{taures}) shows that $\tau$ approaches the 
real axis as $t \to 1$:   At finite $N$, a singularity
at $Im(\tau) = 0$ can be interpreted in terms 
$(p,q)$-branes provided that $\tau$ approaches a rational point on 
the real axis.  In the limit $N \to \infty$  one gets a smooth 
distribution $(p,q)$-branes.  While this $SO(2)$  rotation is
only a symetry for $\lambda =0$, one sees that for $0 < \lambda
<1$ the picture is qualitatively the same, and thus the infra-red 
structure is primarily determined by the flow of the chiral 
multiplet mass.  As we will see, the structure is rather different
for $\lambda=1$.

\subsection{Flows with $\lambda = 1$: Five-branes}

If one looks at (\ref{sbrane}) one sees that
various coefficients either vanish or diverge as 
$\lambda \to 1$.   Re-examining the asymptotics of the 
ten-dimensional metric for $\lambda=1$ yields:
\begin{equation}
ds^2 ~\sim~ \widetilde  \Omega^{1\over 4}~ 
\Big[  2^{2 \over 3}\, 3^{1 \over 3} \, e^{2C_1}\, 
\chi^{2 \over 3} \, \big(\eta_{\mu \nu} dx^\mu dx^\nu 
\big) -  d \chi^2 \Big] ~-~ \widetilde 
\Omega^{-{3\over 4}}~\Big[ ~{1 \over 3 \chi^2 }\, dw_2^2 ~+~  
 d\tilde s_4^2  \Big]  \,,
\end{equation}
where $ d\tilde s_4^2$ is a complicated, but regular metric
on $\IR\IP^3$ and in the $\theta$ direction.  The warp factor,
$\widetilde \Omega$, is now given by:
\begin{equation}  
\widetilde \Omega \equiv  
\coeff{1}{9}\,\big(3\, \cos^2 (2\,\theta )  ~+~ 4\, \sin^2 (2\,\theta) 
\, \sin^2 (\phi)\big)  \,.
\end{equation}

The dilaton matrix takes the form
\begin{equation}
\cS\,\cS^T ~\sim~
\coeff{1}{3}\,\widetilde \Omega^{-{1 \over 2}}~ \left( \matrix{1 + 2 \cos^2
(\theta ) & 2\, \sin(2 \,\theta)\, \cos\phi \cr 2\, \sin(2\, \theta)\, 
\cos\phi & 1 + 2 \sin^2(\theta) }\right)  \,. 
\end{equation}
The metric and the dilaton no longer have a ring singularity,
but only have a singularity at the points 
$\theta = \pm {\pi \over 4}, \phi = 0$.  On the other hand, the
metric now has a singularity at $\chi =0$.  It is not so
simple to give this metric a geometric interpretation,  
particularly since one of the internal directions is blowing up as 
$\chi \to 0$.  Also note that the  coefficient of the $D3$-brane metric 
vanishes as  $\chi \to 0$, which, in principle, suggests that the 
flow probes arbitrarily far into the  infra-red.
 
The metric and dilaton are actually a little more regular near $\chi =0$, 
$\theta = \pm \pi/4$, $\phi =0$, than is apparent from the foregoing.
Setting  $\theta = {\pi \over 4} + \psi$, $t = 1 - {1 \over 2} \chi^2$ 
and expanding in small $\chi, \psi$ and $\phi$ one finds:
\begin{eqnarray}
ds^2 ~\sim~ & 2\, \widetilde  \Omega^{1\over 4}~  \Big[  
3^{1 \over 3} \, e^{2C_1}\,  \chi^{2 \over 3} \, \big(
\eta_{\mu \nu} dx^\mu dx^\nu  \big) -  d \chi^2 \Big] \cr &  ~-~ 
\coeff{1}{2\, \sqrt{3}}  \, \widetilde  \Omega^{-{3\over 4}}~\chi^2~
\Big[ ~ \coeff{16}{3}\, d\psi^2 ~+~   d\phi^2 + \phi^2\, 
(\sigma_1^2 +  \sigma_2^2 +\sigma_3^2 ) ~ \Big]  \,,
\label{fasypmet}
\end{eqnarray}
where one now has:
\begin{equation}  
\widetilde \Omega \equiv  
\coeff{4}{9}\,\big(3\, \psi^2  + \phi^2 + \chi^4 \big) \,.
\end{equation}
Note that the metric (\ref{fasypmet}) has  round 
$\IR \IP^3$ fibers, but there is a conical singularity at $\phi =0$.  
The dilaton matrix becomes:
\begin{equation}  
\cS\,\cS^T  ~\sim~\coeff{2}{3} \,\widetilde 
\Omega^{-{1\over 2}} ~\cQ~ \left( \matrix{2 & -\psi \cr  -\psi &  
2 \psi^2 +\coeff{1}{2} \,\phi^2 + \coeff{1}{2} \,\chi^4} \right)~\cQ^T 
\,, 
\end{equation}
where $\cQ$ is a rotation by $\theta = \pi/4$.
 
Motivated by \cite{PolStr} one would like to interpret this
solution in terms of $5$-branes.  Indeed, it is apparent from the
dilaton configuration that the foregoing solution
appears to represent an S-dual pair of $5$-branes: one of type $(1,1)$ 
and the other of type $(1,-1)$.  The metric contains an $SO(4)$-invariant 
sector which presumably represents the $3$-spheres (or at least $\IR\IP^3$'s)
transverse to the putative $5$-branes.

\section{Supergravity and the phases of super QCD}

The vacua of the $\cN=1^*$ theories have been extensively studied
\cite{OneStar,OneStarNew}.  The basic point is that the chiral multiplet
mass modifies the superpotential so that the ground-state
vevs for the (complex) scalars must satisfy:
\begin{equation}
\big[\Phi_i\,,\Phi_j\big] ~=~ -{m \over  \sqrt{2}} 
\epsilon_{ijk}~ \Phi_k \,.
\label{WVac}
\end{equation}
If the mass parameter, $m$, is real then the solution to this equation
is to take the $\Phi_j$ to be some real combination of the
anti-hermitian generators of an $SU(2)$ subgroup of $SU(N)$.  
It was thus argued in \cite{Myers,PolStr} that the ground states 
of the $\cN=1$ theory should correspond to the $D3$-branes becoming 
dielectric $5$-branes that  wrap a non-commutative $S^2$ whose
radius is determined by  $\sum_j Tr (|\Phi_j|^2 )\sim m^2 C_2(R)$, 
where $C_2(R)$ is the quadratic Casimir of the $SU(2)$ representation,
$R$, induced by the embedding of $SU(2)$ into $SU(N)$.

To relate the field theory to the supergravity flows, recall that for
finite $N$ and for commuting vevs, the $\Phi_j$ may be 
thought of as the Cartesian coordinates transverse to the
$D3$-branes.  More precisely, the solutions here have
an $SO(3)$-invariance:  In (\ref{WVac}) the (real) $SO(3)$ acts 
the indices $i,j,k$, with the real and imaginary parts of $\Phi_i$
transforming separately, each as a triplet of $SO(3)$.  Thus the
real and imaginary parts of $\vec \Phi$ correspond to the 
coordinates $\vec u$ and $\vec v $ on $S^5$.  If we were to obtain
precisely the solution of \cite{PolStr} then
the  $5$-branes should emerge in the limit in which 
$v_j \equiv 0$:  Instead we find (for $\lambda <1$)  a ring singularity  
when $\vec u$ and $\vec v$ are parallel.   The reason for this
difference is easily understood: the supergravity solution has 
an extra continuous duality symmetry, and for $\lambda < 1$
the solution is smeared out into a ``duality averaged'' family
of type $(p,q)$ branes.  This continuous symmetry must be broken
by the string theory, or for finite $N$ field theory, and presumably
the ring singularity  must resolve itself into a discrete  rational family
of type $(p,q)$ branes.   

For $\lambda = 1$, one arrives at something much
closer to the ideas of \cite{Myers,PolStr}.  In particular, the supergravity 
solution has symmetries that are consistent with string theory, and 
finite $N$ field theory, and appear to represent a pair of $S$-dual,
oblique confinement phases of the field theory.

The flows with $\lambda <1$ also exhibit the interesting new feature 
of ``rounding out'' into a $7$-brane world  of  finite scale 
{\it except} when the vevs of $\vec \Phi$ satisfy the
reality condition ($\vec u,\vec v$ parallel) for which
there is a vacuum state in the $\cN=1^*$ theory.   Thus the
supergravity theory ``knows'' that there is no $(3,1)$-dimensional
infra-red, field theory limit below the energy scale of the chiral 
multiplet mass, unless the vevs obey the proper reality condition.

\section{Final comments}

From the detailed analysis of the $\cN=1^*$ flows it is clear 
that working purely in five dimensions conceals much that is
important.  The 
details of how the flows behave in the far infra-red only become
apparent once one lifts to ten dimensions.  Because of the 
warp-factors, apparently singular metrics in five dimensions
become far more regular in ten dimensions, and the singularities
of the ten-dimensional solutions develop a rich but interpretable
structure.  In addition, contrary to naive expectations,
the dilaton and axion flow in the ten-dimensional theory in a
manner that is entirely determined by the running of the masses
of the chiral multiplets and the gaugino vevs.  Apart from giving
information about the running coupling on the brane, the behaviour
of the dilaton and axion also gives important information about the 
types of branes that emerge in the infra-red.   While these details  
are, in principle, encrypted
in the five-dimensional solution, the only way to make them
apparent is to construct the lift to ten dimensions.

The supergravity solutions discussed here  mesh very nicely
with one's expectations based upon field theory.  It is 
interesting to see how the supergravity solution ``rounds-out'' 
so as to prevent access to the infra-red when the scalar
vevs do not satisfy the proper reality conditions.  
One also sees the emergence of various types of dielectric $5$-brane
singularities and, depending on how
the gaugino vev flows, one sees how the symmetries of the
five-dimensional flow act on the $5$-brane sources to
produce a ring singularity, or a discrete pair of singularities.

Another lesson coming from the supergravity solutions described  
here is that the details of the  infra-red singularity can
depend discontinuously upon vevs: Here
the infra-red limit was essentially the same for all $\lambda <1$,
but changed dramatically at $\lambda =1$.  This suggests that 
in the infra-red there are only a  discrete set of choices for
vevs, and that the corresponding choices in the UV dramatically
change the infra-red physics.   This is precisely what is found in the field 
theory \cite{OneStarNew}.   

This points to a draw-back of using ${\cal N}=8$ supergravity:
It requires that one restrict ones attention to a very special 
sub-class of possible excitations.  One has thus made some 
implicit choices of vevs in the UV, and so it will
restrict the class IR fixed point theories.  In particular,
$\cN=8$ supergravity methods can see only a limited subset
of the general family of fixed points  outlined in \cite{PolStr,OneStar,
OneStarNew}.  To access the broader classes of fixed points 
one must allow several different operator vevs to flow independently,
and thus one must go back to the full ten-dimensional theory.

Thus, five-dimensional supergravity methods are 
very powerful in generating exact solutions, whereas competing
ten-dimensional approaches \cite{PolStr} usually have to resort to 
linear approximations.  Moreover, five-dimensional methods
have led to significant insights into  holographic field theories.
However, the five-dimensional methods have limitations:  There are 
restrictions on the families of solutions,  and there is the ``myopia'' 
of the five-dimensional perspective.   It has been shown here
that at least the myopia can be completely corrected by lifting 
to ten dimensions.  

Before concluding I wish to try to extract a further lesson  from
the $\cN=1^*$ flows considered here.  It was shown that the radius
at which the brane-world geometry rounds out is given by
$r_0 \sim \log(m_{UV})$, where $m_{UV}$ is the UV mass parameter
that induces the breaking of $\cN=4$ to $\cN=1$ Yang-Mills.
Such logarithms are very familiar in brane-world scenarios,
and indeed one of the attractive features of these scenarios
is that large energy scale differences are obtained from
small geometric separations precisely through such logarithms.
However, in the supergravity solution presented here, the brane
geometry is not something that is chosen: It is a derived property 
from the ``initial datum,''  $m_{UV}$.  Thus the freely choosable
parameter still suffers from the hierarchy problem, and it raises the
issue as to whether brane-world scenarios actually provide a solution
to the hierarchy problem, or merely yield a geometric 
encoding of a logarithm.

Whatever conclusion one makes about brane-worlds and the hierarchy 
problem, the work described here illustrates a rather more 
general cautionary tale:  To properly
interpret a flow involving brane-world physics one {\it must}
embed the branes in the full, parent string theory

\ack

I would particularly like to thank my collaborator,
Krzysztof Pilch:  The results presented here are
based largely upon our joint work. I am also very grateful to the 
organizers of the G\"ursey Memorial
Conference for
opportunity to present this material, and to the Aspen Center
for Physics for its hospitality while I clarified
some of the issues underlying this work.  This work was supported in part  
by funds provided by the DOE under grant number DE-FG03-84ER-40168.


\section*{References}

\begin{thebibliography}{99}
\small

\bibitem{JMalda}
J. Maldacena,  \nihil{The Large $N$ Limit of Superconformal
Field Theories and Supergravity,}, Adv.~Theor. Math. Phys.~{\bf 2}
(1998) 231; \eprt{hep-th/9711200}.

\bibitem{HolDims}  
S.S.\ Gubser, I.R.\ Klebanov, A.M.\ Polyakov,
\nihil{Gauge Theory  Correlators from Non-Critical String Theory,}
Phys.~Lett.~{\bf B428}  (1998) 105, \eprt{hep-th/9802109}; 
\newline
E.\ Witten, \nihil{Anti-de Sitter Space and Holography,}
Adv. Theor.  Math.  Phys. {\bf 2} (1998) 253, \eprt{hep-th/9802150}. 
 
\bibitem{flowpaps}
L.\ Girardello, M.\ Petrini, M.\ Porrati and
A.\ Zaffaroni,  \nihil{Novel Local CFT and Exact Results on
Perturbations of $N=4$ Super Yang--Mills from AdS Dynamics,}  
\jhep{12} (1998) 022, \eprt{hep-th/9810126}; 
\newline
J.\ Distler and F.\ Zamora,
\nihil{Nonsupersymmetric Conformal Field Theories from Stable Anti-de
Sitter Spaces,}  \atmp{2} (1999) 1405, \eprt{hep-th/9810206}. 

\bibitem{FGPWa}
 D.~Z. Freedman, S.~S. Gubser, K.~Pilch, and N.~P. Warner,
\nihil{Renormalization Group Flows from Holography---Supersymmetry
and a c-Theorem,} CERN-TH-99-86, ; to appear in \atmp{3} (2000);
\eprt{hep-th/9904017}.

\bibitem{NSVZ}
V.~Novikov, M.~A.\ Shifman, A.~I.\ Vainshtein, V.~Zakharov, \nihil{Exact
Gell-Mann-Low Function of Supersymmetric Yang-Mills Theories from
Instanton Calculus,} Nucl. Phys. {\bf B229} (1983) 381; \hfil\break 
I.~I.\ Kogan, M.~A.\ Shifman, and A.~I.\ Vainshtein, \nihil{Matching
Conditions and Duality in N=1 SUSY Gauge Theories in the Conformal
Window,} Phys. Rev.~{\bf D53} (1996) 4526, \eprt{hep-th/9507170}.  
 
\bibitem{KPW}
A.~Khavaev, K.~Pilch and N.P.~Warner,
\nihil{New Vacua of Gauged N=8 Supergravity in Five Dimensions,}
\plt{487} (2000) 14;  \eprt{hep-th/9812035}. 

\bibitem{FGPWb}
D.~Z. Freedman, S.~S. Gubser, K.~Pilch, and N.~P. Warner,
{\it Continuous Distribution of D3-branes and Gauged Supergravity,}
 \jhep{7} (2000) 38;  \eprt{hep-th/9906194}.  
 
\bibitem{GPPZ}
L. Girardello, M. Petrini, M. Porrati and A.
Zaffaroni \nihil{The Supergravity Dual of $\cN = 1$ Super-Yang-Mills 
Theory,} \nup{569} (2000) 451, \eprt{hep-th/9909047}. 
 
\bibitem{Beh}
K.~Behrndt,
\nihil{Domain walls of D = 5 supergravity and Fixed Points of $\cN = 1$ 
Super Yang-Mills}, \eprt{hep-th/9907070}. 
 
\bibitem{PetZaf}
M.~Petrini and A.~Zaffaroni,
\nihil{The Holographic RG Flow to Conformal and Non-Conformal Theory},
\eprt{hep-th/0002172}. 
 
\bibitem{NEMP} 
N.~Evans and M.~Petrini, \nihil{AdS
RG Flow and the Super-Yang-Mills Cascade,}  SHEP-00-05,
IMPERIAL-TP-99-00-28,  \eprt{hep-th/0006048}. 

\bibitem{PolStr}
J.~Polchinski and M.~J.~Strassler,
{\it The String Dual of a Confining Four-Dimensional Gauge Theory,}
\eprt{hep-th/0003136}. 

\bibitem{Myers}
R.C.~Myers, \nihil{Di-electric Branes}, \jhep{9912}  (1999) 22;
\eprt{hep-th/9910053}.

\bibitem{KPNWa} K.~Pilch and N.P.~Warner,
\nihil{ A New Supersymmetric Compactification of Chiral, IIB 
Supergravity,}    CITUSC/00-012, USC-00/01;
\plt{487}  (2000) 22; \eprt{hep-th/0002192}. 
 
\bibitem{KPNWb}
 K.~Pilch and N.P.~Warner,
\nihil{${\cal N}\!=\!2$ Supersymmetric RG Flows and the IIB 
Dilaton,}  CITUSC/00-18, USC-00/02; \eprt{hep-th/0004063}. 
 
\bibitem{KPNWc}
 K.~Pilch and N.P.~Warner,
\nihil{${\cal N}=1$ Supersymmetric Renormalization 
Group Flows from IIB Supergravity,}    CITUSC/00-024, USC-00/03,
 \eprt{hep-th/0006066}. 
 
\bibitem{RS}
 L.~Randall and R.~Sundrum, \nihil{A Large
Mass Hiearchy from a Small Extra Dimension,} \prl{83} (1999) 3370,
\eprt{hep-ph/9905221}; \nihil{An Alternative to Compactification,} 
\prl{83} (1999) 4690, \eprt{hep-th/9906064}; 

\bibitem{BVV}
J. de Boer, E.~Verlinde and H.~Verlinde,
\nihil{On the Holographic Renormalization Group,}
\jhep{8} (2000) 3; \eprt{hep-th/9912012}.

\bibitem{GaugSG}
M.\ G\"unaydin, L.J.\ Romans and N.P.\ Warner,
\nihil{Gauged $N=8$ Supergravity in Five Dimensions,}
Phys.~Lett.~{\bf 154B} (1985) 268; \nihil{
Compact and Non-Compact Gauged Supergravity Theories in Five Dimensions,}
Nucl.~Phys.~{\bf B272} (1986) 598. \hfil \break
M.~Pernici, K.~Pilch and P. van Nieuwenhuizen,
\nihil{Gauged $N=8$ $D = 5$ Supergravity,}
Nucl.~Phys.~{\bf B259} (1985) 460. 

\bibitem{MetAns}
B.\ de Wit and H.\ Nicolai, \nihil{On the Relation
Between $d=4$ and $d=11$ Supergravity,} Nucl.~Phys.~{\bf B243}
(1984) 91; \hfil \break
B.\ de Wit, H.\ Nicolai and N.P.\ Warner,
\nihil{The Embedding of Gauged $N=8$ Supergravity into $d=11$
Supergravity,}
Nucl.~Phys.~{\bf B255} (1985) 29.
 
\bibitem{OneStar}
C.~Vafa and E.~Witten,
\nihil{A Strong Coupling Test of S Duality,}
Nucl.~Phys.~{\bf B431} (1994) 3, \eprt{hep-th/9408074};
\newline
R.~Donagi and E.~Witten, \nihil{Supersymmetric Yang-Mills 
Theory and Integrable Systems,}
Nucl.~Phys.~{\bf B460} (1996) 299; \eprt{hep-th/9510101}.

\bibitem{OneStarNew}
N.~Dorey,  \nihil{An Elliptic Superpotential for Softly 
Broken ${\cal N}=4$ Supersymmetric Yang-Mills Theory,}  \jhep{9907} (1999) 
021; \eprt{hep-th/9906011}. 
\newline
N.~Dorey and S.~Prem~Kumar,  \nihil{Softly-Broken ${\cal N} = 4$ Supersymmetry 
in the Large-N Limit,} \jhep{0002} (2000) 006, \eprt{hep-th/0001103};
\newline  
O.~ Aharony, N.~Dorey and S.~Prem~Kumar,
\nihil{New Modular Invariance in the ${\cal N} =1^*$ Theory, Operator 
Mixings and Supergravity Singularities,} \eprt{hep-th/0006008}. 
 
\end{thebibliography}
\end{document}